\documentclass[conference]{IEEEtran}
\usepackage{amsmath,amssymb,graphicx,booktabs}
\usepackage{pgfplots}
\pgfplotsset{compat=1.18}
\usepgfplotslibrary{groupplots}
\usepackage[inline]{enumitem}
\usepackage{algorithm}
\usepackage{algpseudocode}
\usepackage{placeins} 
\title{Study of Graph-Based Search for Energy-Efficient Clustering in Cell-Free Massive MIMO Networks}
\author{Julio Cesar Cardoso Tesolin and Rodrigo C. de Lamare\\
Centre for Telecommunications Studies, PUC-Rio, Brazil\\
Emails: jcctesolin@gmail.com, delamare@puc-rio.br}
\begin{document}
\maketitle
\begin{abstract}
This paper investigates energy-efficient clustering in user-centric cell-free massive MIMO networks, addressing the access point clustering and power allocation problems via a mixed-integer fractional program. We propose a framework for energy-efficient clustering and power allocation with a graph-based structured search and describe its optimum solution via an exhaustive search. We also develop the Graph-Based Steepest Ascent (GBSA) algorithm, which combines a graph-based structured search along with continuous power allocation via fractional programming. The proposed GBSA algorithm achieves linear per-iteration complexity while reaching energy efficiency close to the global optimum, outperforming competing techniques and offering a scalable solution for future networks.
\end{abstract}
\section{Introduction}
\IEEEPARstart{U}{ser}-centric (UC) cell-free (CF) massive multiple-input multiple-output (mMIMO) systems \cite{2021:ozlem:foundations,mmimo,wence} has become a key architecture for next-generation wireless systems. By replacing fixed cells with many distributed Access Points (APs), CF-mMIMO significantly improves uniformity of service and spectral efficiency. In the UC approach, users are served only by APs with strong large-scale fading coefficients, reducing fronthaul load and improving scalability \cite{2021:ozlem:foundations,cfrmmse,itaps}. While this cooperative model enhances quality of service, 
it also increases power consumption, making energy efficiency (EE) a critical requirement. Optimization strategies are therefore essential to balance high service quality with energy expenditure control.

Prior efforts have investigated UC clustering to reduce signaling and computational burden \cite{2021:gennaro:rnn}, \cite{2022:mashdour:fairness}, \cite{2022:mashdour:enhancedgreedy}, \cite{rscf}. Growing attention has been given to resource allocation and energy-driven UC clustering \cite{2023:ozlem:oran}, \cite{2022:tan:hierarchical}, \cite{2023:mendoza:drm},\cite{rra} often employing complex schemes like deep reinforcement learning to achieve effective performance–power trade-offs \cite{2023:shi:risdrml}. Despite these advancements, the combinatorial complexity of the EE-optimal cluster selection problem remains a significant challenge, demanding scalable and tractable solutions that avoid the high cost of exhaustive search.

In this work, we investigate energy-efficient cluster selection and power allocation in UC CF-mMIMO downlink networks. We present a framework for energy-efficient clustering and power allocation with a graph-based structured search and describe its optimum solution via an exhaustive search \cite{mio}. We also devise the Graph-Based Steepest Ascent (GBSA) algorithm, which combines a graph-based structured search along with continuous power allocation via fractional programming. The GBSA algorithm uses graph-based modeling and fractional programming to maximize radio access network (RAN) energy efficiency. This approach ensures scalability and tractability, with simulations showing performance superior to competing techniques and close to the global optimum.

{\it Notation}: Scalars are denoted by $a$, $A$; column vectors by $\mathbf{a} \in \mathbb{C}^n$; and matrices by $\mathbf{A} \in \mathbb{C}^{m \times n}$. $\mathcal{A}$ denotes a set. $(\cdot)^T$ is the transpose, $\mathbb{E}\{\cdot\}$ is the statistical expectation, and $\#(\cdot)$ is the set cardinality. $\mathcal{O}(\cdot)$ represents the worst-case asymptotic complexity.
\section{System Model}
Energy consumption has become a critical performance factor in wireless networks that enable cooperation across distributed APs, allowing for highly flexible and scalable deployments. In this context, energy efficiency (EE) measures the trade-off between the network utility function and the total power expenditure required to sustain communication and signal processing operations \cite{2015:zappone:eefractional}. A general form of EE metric can be expressed as
\begin{equation}
\mathrm{EE} = \frac{f(\mathrm{SINR})}{g\left( \text{Power Consumption} \right)},
\label{eq:ee}
\end{equation}
where $f(\cdot)$ is a non-negative utility function (e.g., achievable rate), and $g(\cdot)$ is a positive, monotonically increasing function modeling the total power consumption.

Let us consider a UC CF-mMIMO network \cite{c&didd,iddllr,iddocl,ccaps} modeled using the dynamic cooperation clustering (DCC) framework proposed in \cite{2021:ozlem:foundations}. It operates under time-division duplexing (TDD) and orthogonal frequency-division multiplexing (OFDM). The network consists of $L$ APs connected to a central processing unit (CPU) with unlimited fronthaul bandwidth. Each AP is equipped with $N$ antennas, which cooperatively serve $K$ single-antenna UEs. Due to the envisioned ultra-dense characteristics, we assume that the aggregate number of antennas in the network  significantly exceeds the total number of UEs, i.e., $NL \gg K$ \cite{2021:ozlem:foundations}. 
To emphasize the core aspects of distributed cooperation and signal processing, we assume that 
\begin{enumerate*}[label=(\roman*)]
    \item perfect channel state information is available at the transmitters;
    \item APs and UEs are synchronized in time and frequency;
    \item the system operates in narrowband or per-subcarrier OFDM mode;
    \item hardware impairments 
    are neglected; and
    \item  the channel is modeled as a linear combination of channel components with additive white Gaussian noise (AWGN).
\end{enumerate*}
\begin{figure}[!htb]
	\centering
	\includegraphics[width=0.5\linewidth]{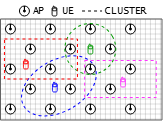}
    \caption{User Centric CF-mMIMO Network}
	\label{fig:uccfmimo}
\end{figure}

In this sense, let $s_k \in \mathbb{C}$ denote the data symbol intended for UE $k \in \mathcal{K}=\{1, \dots, K\}$, with normalized power $\mathbb{E}\{|s_k|^2\} = 1$, and let $\mathbf{w}_{k,l} = \begin{bmatrix}w_{k,l,1} & \dots & w_{k,l,N} \end{bmatrix}^T \in \mathbb{C}^{N}$ be the normalized precoding vector at AP $l\in \mathcal{L}=\{1, \dots, L\}$ for UE $k$, such that $\|\mathbf{w}_{k,l}\|^2 = 1$. Let us also define a binary variable $c_{k,l} \in \{0,1\}$ to indicate whether AP $l$ participates in the transmission to UE $k$.
Consequently, let $\mathbf{C} = [c_{k,l}] \in \{0,1\}^{K \times L}$ denote the matrix of the binary association variables, representing a given RAN serving state where $\sum_{l=1}^L c_{k,l} \ge 1$.
Considering that each AP typically has a few antennas connected to a single power amplifier, 
let $\sqrt{p_{k,l}} \in \mathbb{R}_+$ denote the per-AP power scaling variable to manage the assigned power for UE $k$ at AP $l$, and $\mathbf{P} = [p_{k,l}] \in \mathbb{R}_+^{K \times L}$ denote a given RAN power state. Therefore, the precoded signal transmitted by AP $l$ to UE $k$ is given by $\mathbf{x}_{k,l} = \sqrt{p_{k,l}}\, c_{k,l}\, \mathbf{w}_{k,l} s_k,$ and the total transmit signal from AP $l$ to all UEs is given by $\mathbf{x}_{l} = \sum_{k=1}^K \sqrt{p_{k,l}}\, c_{k,l}\, \mathbf{w}_{k,l} s_k \in \mathbb{C}^{N}$, 
and the total power transmitted by AP $l$ is given by
$p_l = \mathbb{E}\left\{\|\mathbf{x}_l\|^2\right\} 
= \sum_{k=1}^K p_{k,l}c_{k,l}$, restricted by $p_{l}^{\min} \leq p_{l} \leq p_{l}^{\max}$.
The transmitted signals propagate over a wireless medium, and the channel vector between the $n$-th antenna of AP $l$ and UE $k$ is denoted by $\mathbf{g}_{k,l} = \begin{bmatrix}g_{k,l,1} & \dots & g_{k,l,N}\end{bmatrix}^T \in \mathbb{C}^{N}$, modeled as $\mathbf{g}_{k,l} = \sqrt{\beta_{k,l}} \cdot \mathbf{h}_{k,l}$ , where $\beta_{k,l} \in \mathbb{R}_+$ is the large-scale fading coefficient (accounting for path loss and shadowing), and $\mathbf{h}_{k,l} \in \mathbb{C}^{N}$ is the small-scale fading vector. Thus, the total received signal at UE $k$ is described by
{\small
\begin{align}
& y_k = 
\sum_{l=1}^L  \mathbf{g}_{k,l}^T \left(\sum_{i=1}^K \sqrt{p_{i,l}}\, c_{i,l}\, \mathbf{w}_{i,l} s_i \right) + n_k = 
\\ \notag
& \underbrace{\sum_{l=1}^L \mathbf{g}_{k,l}^T\, \mathbf{w}_{k,l}\, \sqrt{p_{k,l}}\, c_{k,l} s_k}_{\text{desired signal: } y_{k}^\mathrm{S}} +
\underbrace{\sum_{l=1}^L \mathbf{g}_{k,l}^T \left( \sum_{\substack{i=1 \\ i\neq k}}^K \sqrt{p_{i,l}}\, c_{i,l}\, \mathbf{w}_{i,l} s_i \right)}_{\text{interference: } y_{k}^\mathrm{I}} + n_k,
\end{align}
}where $y_k^{\mathrm{S}}$ and $y_k^{\mathrm{I}}$ are, respectively, the desired signal and interference at UE $k$, and $n_k \sim \mathcal{CN}(0, \sigma^2)$ is the AWGN. The resulting signal-to-interference-plus-noise ratio (SINR) at UE $k$ is 
\begin{equation}
\mathrm{SINR}_k = 
\frac{
\mathbb{E}\left\{ \left| y_k^{\mathrm{S}} \right|^2 \right\}
}{
\mathbb{E}\left\{ \left| y_k^{\mathrm{I}} \right|^2 \right\} + \sigma^2
},
\label{eq:sinrk}
\end{equation}
and the achievable downlink data rate can be expressed as 
\begin{equation}
R_{k}=B\log_{2}{(1+\mathrm{SINR}_{k})}
\label{eq:dwthrp}
\end{equation}
where $B$ is the total downlink bandwidth.

Without loss of generality, we focus on the RAN energy efficiency (RANEE), using the aggregate achievable downlink data rate as the network utility function. Thus, using (\ref{eq:ee},\ref{eq:sinrk},\ref{eq:dwthrp}), RANEE can be defined as
\begin{equation}
\mathrm{RANEE}(\mathbf{P},\mathbf{C}) = 
\frac{\sum_{k=1}^K R_k(\mathbf{P},\mathbf{C})}{LP_{\mathrm{circuit}}+\sum_{l=1}^L p_{l}(\mathbf{P},\mathbf{C})}.
\label{eq:ree}  
\end{equation}
The numerator is given in bits per second, while the denominator captures the total energy consumption in watts, comprising the sum of transmit powers $p_{l}$ over all APs in the network, and the constant term $P_{\mathrm{circuit}}$, assumed equal for all APs. This last term accounts for the static power consumed by baseband processing, fronthaul communication, synchronization, and cooling systems at the APs. Note that EE is a non-monotonic function of the transmit power budget. While higher transmit power typically improves $\text{SINR}$-based metrics, EE often attains its maximum at intermediate power levels, where the marginal utility gain is balanced by the corresponding increase in power expenditure  \cite{2015:zappone:eefractional}.

\section{Graph Modeling of the User-Centric Clustering Universe}\label{sec:ranstate}

Following the DCC approach, let $\mathcal{D}_k \subseteq \mathcal{L}$ denote the non-empty set of APs eligible to serve UE $k$. These candidates may be selected based on physical-layer constraints (e.g., channel gain, pilot contamination), network-layer limitations (e.g., hardware constraints), or whether AP $l$ positively contributes to a performance metric when serving UE $k$. The association of APs to UE $k$ is captured by the binary decision variable $c_{k,l} \in \{0,1\}$. Let $\mathcal{S}_k$ be the set of all possible AP subsets able to serve UE $k$. Considering $D_k=\#(\mathcal{D}_k)$, the cardinality of $\mathcal{S}_k$ is ${S}_k =\#(\mathcal{S}_k) = 2^{D_k} - 1$.

A RAN serving state is modeled by a bipartite graph $\mathtt{B}=(\mathcal{K},\mathcal{L},\mathcal{E})$, where the edge set $\mathcal{E}$ represents the AP-UE associations \cite{ldpc,memd}. The entire set of all possible RAN serving states $\mathcal{U}$ is obtained by selecting one serving set $\mathcal{S}_{k,i}$ for each UE $k$. The total number of unique serving states $U$ is:
\begin{equation}
U=\#(\mathcal{U}) = \prod_{k=1}^{K}S_k= \prod_{k=1}^{K} \left(2^{{D}_k} - 1\right).
\label{eq:gclustercount}
\end{equation}
Due to the envisioned increase in $K$ and $L$, $U$ rapidly escalates, ranging from the lower bound (single-AP service) to the upper bound (full cell-free cooperation): $1 \leq U \leq (2^{L} - 1)^K$. This combinatorial explosion highlights the need for scalable search algorithms to avoid computationally prohibitive exhaustive searches.
Thus, to structure the search space $\mathcal{U}$, we propose to model the RAN serving states as an undirected graph $\mathtt{R}=(\mathcal{U},\mathcal{E})$. In this model, each state $\mathtt{B} \in \mathcal{U}$ is a vertex represented by its adjacency matrix $\mathbf{C} = [c_{k,l}] \in \{0,1\}^{K \times L}$, is an edge connecting two distinct adjacent nodes ($\mathcal{E} \subseteq \{\mathcal{U} \times \mathcal{U}\}$ , and $\phi$ is the incident function that maps each edge to the pair of vertices it connects($\phi:\mathcal{E} \rightarrow \mathcal{U} \times \mathcal{U}$). Given the binary nature of matrix $\textbf{C}$, we employ the Hamming distance as the adjacency criterion to structure the graph $\mathtt{R}$.  Hence, Hamming distance ($d_H$) between their adjacency matrices $\mathbf{C}$ and $\mathbf{C}'$ is defined as:
\begin{equation}
d_H(\mathbf{C}, \mathbf{C}') = \sum_{k=1}^{K}\sum_{l=1}^{L} \mathbf{1}_{\{c_{k,l} \neq c_{k,l}^{\prime}\}}.
\label{eq:Hdist}
\end{equation}
Moreover, two adjacent states $\mathtt{B}$ and $\mathtt{B}'$ are connected if their Hamming distance is at most a threshold $m$: $\phi(\mathtt{B},\mathtt{B}^{\prime}) = 1$ if $d_H(\mathbf{C},\mathbf{C}^{\prime}) \leq m$, and 0 otherwise. This criterion ensures that the graph $\mathtt{R}$ is a regular graph and defined by the number of configurations differing by at most $m$ bits:
\begin{equation}
\deg_m(\mathbf{C}) = \sum_{i=1}^{m} \binom{KL}{i}.
\label{eq:dgrC} 
\end{equation}

\section{Proposed EE Search-and-Evaluate Approach}\label{sec:scheval}

A central challenge in UC CF-mMIMO is to carry resourc allocation \cite{tds,tds2,rapa} and ensure each UE’s QoS while minimizing total power consumption. Each AP must choose transmit powers within $(p^{\min}_l, p^{\max}_l)$ and ensure every UE meets the rate target $R_{\min}$. The optimization jointly determines the powers $p_{k,l}$ and cooperation variables $c_{k,l}$ that maximize the RANEE in \eqref{eq:ree}, under power and rate constraints. This yields a mixed-integer nonlinear fractional problem involving discrete clustering, fractional SINR terms, and a nonconvex rate–power ratio.
Among the existing methods for handling fractional programs, Dinkelbach’s algorithm \cite{2015:zappone:eefractional} remains the most widely adopted because it transforms the ratio objective into a sequence of tractable subproblems with subtractive objectives:
\begin{multline*}
    \max_{x\in\mathcal{X}}\frac{f(x)}{g(x)}
\ \Longleftrightarrow
\text{solve iteratively } \max_{x\in\mathcal{X}} (f(x)-\lambda g(x)),
\end{multline*}
where $\lambda^{(t+1)}=f(x^{(t)})/g(x^{(t)})$ is updated until the improvement falls below a tolerance $\epsilon$. When $f(\cdot)$ is concave, $g(\cdot)$ is positive and convex, and $\mathcal{X}$ is convex, this procedure converges super-linearly to the global optimum \cite{2015:zappone:eefractional}\cite{2015:zappone:eepwrctrl}.

However, the introduction of binary cooperation variables $\mathbf{C}$ renders the feasible set nonconvex and combinatorial. In this mixed-integer fractional case, direct fractional programming cannot be applied, and two main strategies arise: 
\begin{enumerate*}[label=(\roman*)]
    \item continuous relaxation followed by rounding, or
    \item decomposition between discrete and continuous components \cite{2015:zappone:eefractional},\cite{2015:zappone:eepwrctrl}
\end{enumerate*}. 
Relaxation often increases computational burden and may produce infeasible rounded solutions, especially under coupled SINR constraints.

\subsection{The Search-and-Evaluate Approach}
To mitigate the limitations inherent to the relaxation-and-rounding approach, we adopt a decomposition-based search-and-evaluate approach, where binary and continuous variables are optimized separately but coherently. Thus, the RANEE maximization can be written as
\begin{equation}
\begin{aligned}
\max_{\substack{\mathbf{C}\in\{0,1\}^{K\times L} \\ \mathbf{P}\ge0}}
& \text{RANEE}(\mathbf{P},\mathbf{C})
= \frac{\sum_{k=1}^{K} R_k(\mathbf{P},\mathbf{C})}
{L P_{\mathrm{circuit}}
+\sum_{l=1}^{L}\sum_{k=1}^{K} c_{k,l} p_{k,l}}
\\
& \text{subject to} \\
& R_k(\mathbf{P},\mathbf{C})\ge R_{\min},\ \forall k,\\
& p^{\min}_l\le \sum_{k=1}^{K} c_{k,l}p_{k,l}\le p^{\max}_l,\ \forall l.
\end{aligned}
\end{equation}
The binary matrix $\mathbf{C}$ defines the RAN serving state, while the continuous matrix $\mathbf{P}$ represents the power allocation. For any fixed $\mathbf{C}$, the inner continuous subproblem
\begin{equation}
\mathbf{P}^*(\mathbf{C})
=\arg\max_{\mathbf{P}> 0} \quad \text{RANEE}(\mathbf{P},\mathbf{C})
\end{equation}
is a pseudoconcave fractional program that can be efficiently solved using Dinkelbach’s algorithm. The outer combinatorial search
\begin{equation}
\mathbf{C}^*
=\arg\max_{\mathbf{C}\in\mathcal{C}}\quad \text{RANEE}(\mathbf{P}^*(\mathbf{C}),\mathbf{C})
\end{equation}
is then performed over feasible serving states modeled as graph vertices. This two-layer decomposition enables different graph-based strategies to be embedded while preserving the exact fractional optimization for each candidate configuration.

Recent studies adopted similar hierarchical decompositions in energy-efficient cell-free systems. In \cite{2022:tan:hierarchical} and \cite{2023:mendoza:drm}, deep reinforcement learning is used to approximate the outer search, while \cite{2023:shi:risdrml} applies multi-agent learning for joint AP selection and precoding. Building on this rationale, we propose a deterministic search stage based on graph topology modeling of the RAN serving states, ensuring tractability and interpretability.

\subsection{Graph-Based Searching Algorithms}
The graph representation of serving states (Section \ref{sec:ranstate}) provides an intuitive structure for discrete search. While exhaustive strategies like BFS or DFS guarantee optimality, they scale exponentially, making them infeasible for large networks. Informed searches such as A* lack admissible heuristics for non-convex fractional objectives, while stochastic metaheuristics, though capable of escaping local optima, incur high computational costs due to numerous RANEE evaluations \cite{2016:russell:aimodernapproach,2023:Zhang:hcimprove}. In contrast, Hill-Climbing (HC) algorithms offer a balanced compromise between exploration and efficiency, exploring memorization to avoid recomputing values \cite{2023:Zhang:hcimprove}. Specifically, the Steepest-Ascent HC (SAHC) variant, which evaluates all neighbors within a Hamming radius $m$ and selects the best one, ensures monotonic objective function improvement with predictable complexity \cite{2016:russell:aimodernapproach,2023:Zhang:hcimprove}. The computational behavior of SAHC is governed by the iteration time ($t_{\text{iter}}$) and the accumulated runtime until convergence ($t_{\text{acum}}$) given by $t_{\text{iter}} = (t_{\text{adjc}} + t_{\text{EE}}) N_{\text{eval}}$ and $t_{\text{acum}} = t_{\text{iter}} N_{\text{mov}}$. The attribute  $t_{\text{adjc}}$ is the state generation time, $t_{\text{EE}}$ is the Dinkelbach optimization time, $N_{\text{eval}}$ is the number of evaluations per iteration, and $N_{\text{mov}}$ is the number of state transitions.

The time required to generate and return an adjacent matrix $\mathbf{C}^{\prime}$ $t_{adjc}$ scales with the binary matrix dimension $K \times L$, leading to $\mathcal{O}(t_{\text{adjc}}) = \mathcal{O}(KL) = \mathcal{O}(n)$, where $n = KL$. The optimization and RANEE computation $t_{EE}$ involve matrix operations within Dinkelbach’s method. Considering that each iteration of Dinkelbach’s algorithm solves a convex subproblem involving up to $n$ continuous variables, its worst-case computational cost is $\mathcal{O}(n^{3})$ \cite{2015:zappone:eefractional}\cite{2015:zappone:eepwrctrl}. Also, each node has $\deg_{m}(\mathbf{C})$ neighbors, resulting in $N_{eval} = \deg_{m}(\mathbf{C})$. Thus, the per-iteration complexity is $\mathcal{O}(t_{\text{iter}})= \mathcal{O}((t_{\text{adjc}} + t_{\text{EE}}) N_{\text{eval}})= \mathcal{O}(n^{3} \deg_{m}\mathbf{C}))$.  Assuming convergence after visiting all vertices, the number of transitions satisfies $N_{mov} = (U - 1)/N_{eval} = (U - 1)/\deg_{m}(\mathbf{C})$.  Also, homogeneous user cooperation limits $D_k = D, \forall k$, leading to $U = (2^{D} - 1)^{K}$. Thus, the accumulated runtime yields $\mathcal{O}(t_{\text{acum}})= \mathcal{O}(n^{3}[(2^{D} - 1)^{K} - 1]) \simeq \mathcal{O}(n^{3} 2^{DK}).$

This analysis reveals that while the accumulated complexity bound is topologically invariant, the iteration complexity grows combinatorially with $m$. Consequently, the minimal-distance case ($m = 1$) minimizes the computational burden per step ($\deg_{1}(\mathbf{C}) = n$), offering the most favorable trade-off between runtime and the ability to escape poor local optima.

\section{Proposed Graph-Based Steepest Ascent Algorithm }
In this section, we detail the Graph-Based Steepest Ascent (GBSA) algorithm. GBSA employs graph modeling to regulate the search space, where the neighborhood of a RAN state $\mathbf{C}$ is defined by the minimal Hamming distance, i.e.,
\begin{equation}
\mathcal{N}(\mathbf{C}) = { \mathbf{C}^{\prime} \in \mathcal{U} : d_{H}(\mathbf{C},\mathbf{C}^{\prime})=1 }.
\label{eq:nghdef}
\end{equation}
where
$d_H(\mathbf{C}, \mathbf{C}') = \sum_{k=1}^{K}\sum_{l=1}^{L} \mathbf{1}_{\{c_{k,l} \neq c_{k,l}^{\prime}\}}$.
The algorithm initializes by retrieving the existing precoders and power allocation $\mathbf{P}=[p_{k,l}]$, from which the initial state $\mathbf{C}$ is derived by setting $c_{k,l} = 1$ if $p_{k,l} > 0$, and 0 otherwise. Once matrices $\mathbf{P}$ and $\mathbf{C}$ are built, the energy efficiency of the current state is calculated using 
\begin{equation*}
\mathrm{RANEE}(\mathbf{P},\mathbf{C}) = 
\frac{\sum_{k=1}^K R_k(\mathbf{P},\mathbf{C})}{LP_{\mathrm{circuit}}+\sum_{l=1}^L p_{l}(\mathbf{P},\mathbf{C})}.
\end{equation*}
as in (\ref{eq:ree}). 
To ensure scalability, we prune the search space by incorporating network topological knowledge. We hypothesize that potential servers for a UE $k$ are limited to the physical neighbors of the APs currently serving that UE. We construct a masking matrix $\mathbf{M} \in \{0,1\}^{K\times L}$, where $\mathbf{M}(k,l)=1$ indicates a valid flip. This is derived by:
\begin{enumerate*}[label=(\roman*)]
\item identifying APs currently serving UE $k$; and
\item marking their registered physical neighbor APs as eligible candidates.
\end{enumerate*}
The set of adjacent states is then generated via bitwise XOR operations with elementary matrices $\mathbf{E}^{(k,l)}$ (where $\mathbf{E}^{(k,l)}_{k,l}=1$) strictly for valid positions $\mathbf{C}^{\prime} = \mathbf{C} \oplus \mathbf{E}^{(k,l)}, \forall (k,l) : \mathbf{M}(k,l)=1.$

For every candidate state $\mathbf{C}^{\prime}$, the optimal power allocation $\mathbf{P}^{\prime}$ is obtained via Dinkelbach’s algorithm. This iteratively solves the concave subproblem:
\begin{equation}
\mathbf{P}^* = \arg \max_{\mathbf{P}} \left[ \sum_{k=1}^K R_k(\mathbf{P},\mathbf{C}^{\prime}) - \lambda^{(t)} P_{\text{tot}}(\mathbf{P}) \right],
\end{equation}
updating $\lambda^{(t)}$ until convergence ($\leq \epsilon_{\text{dinkel}}$), where $P_{\text{tot}}$ is the denominator of \eqref{eq:ree}. The neighbor yielding the highest RANEE is selected. If this candidate improves the current EE by a margin $\epsilon_{\text{EE}}$, the state is updated; otherwise, the search terminates. The procedure is bounded by $N_{\text{mov}}^{\max}$ to prevent excessive runtime, as summarized in Algorithm \ref{algo:gbsaalgo}.

\algnewcommand\algorithmicforeach{\textbf{for each}}
\algdef{S}[FOR]{ForEach}[1]{\algorithmicforeach\ #1\ \algorithmicdo}

\begin{algorithm}
\caption{The proposed GBSA Algorithm} 
\label{algo:gbsaalgo}
\begin{algorithmic}[1]
\Require Initial \textbf{C}, \textbf{P}, neighbor lists, $\epsilon_{EE}$, $N_{mov}^{max}$
\Ensure $\mathbf{C} \not= 0,\mathbf{P} \not= 0$
\State $\mathbf{C} \gets \mathbf{C}_0$, $\mathbf{P} \gets \mathbf{P}_0$
$N_{\text{mov}} \gets 0$\;
\State \Call{RANEE}{\textbf{P,C}}
\State $\mathbf{M} \gets$ \Call{buildMask}{\textbf{C}}
\While{$N_{mov} < N_{mov}^{max}$}
    \State $\mathcal{N}(\mathbf{C}) \gets$ \Call{genAdj}{$\mathbf{C},1$}
    \ForAll{$\mathbf{C^{\prime}} \in \mathcal{N}(\mathbf{C})$)}
    \State $\mathbf{C^{\prime}} \gets$ \Call{applyMask}{$\mathbf{C^{\prime}}$,\textbf{M}}
    \State \Call {RANEE}{$\mathbf{P^{\prime}},\mathbf{C^{\prime}}$}
    \EndFor   
    \State $\textbf{P}^*,\textbf{C}^* \gets \arg\max \text{RANEE}(\mathbf{P^{\prime}},\mathbf{C^{\prime}})$

    \If {$\text{RANEE}(\mathbf{P}^*,\mathbf{C}^*) - \text{RANEE}(\mathbf{P},\mathbf{C}) >\epsilon_{EE}$}
        \State $\textbf{C} \gets \textbf{C}^*$\
        \State recalculate $\mathbf{P}$ and $\mathbf{P^*}$
        \State $N_{mov} \gets N_{mov} + 1$
    \Else
        \State \textbf{break}
    \EndIf
\EndWhile
\State $\mathbf{C} \gets \mathbf{C^*}$, $\mathbf{P} \gets \mathbf{P^*}$
\end{algorithmic}
\end{algorithm}
\vspace{-2mm}

\section{Numerical Results}
This section provides a detailed assessment of the proposed GBSA algorithm to verify its effectiveness in forming user-centric (UC) clusters based on RAN energy efficiency. The evaluation compares GBSA against two reference strategies: (i) an exhaustive search, which delivers the global optimal RANEE and therefore serves as a performance benchmark, and (ii) a joint-optimization (JO) method that applies continuous relaxation followed by a rounding procedure, as described in Section \ref{sec:scheval}.

The analysis focuses on two metrics: the total processing time required to obtain the highest RANEE and the sub-optimality gap relative to the global optimum produced by the exhaustive search. Because the exhaustive approach is computationally demanding, simulations are conducted in a small-scale UC CF-mMIMO scenario comprising 6 APs equipped with 2×2 MIMO antennas and serving up to 5 UEs. The APs are uniformly placed within a 200 × 200 m dense-urban region, following the radio propagation assumptions specified in 3GPP TS 38.901 \cite{2025:3gpp:38901}. While the AP locations and precoding vectors (MRT) remain fixed, UE positions are randomly generated within the service area. For each run, a random deployment seed and a maximum UC cluster size are selected.

Figure \ref{fig:plotComp} evaluates the behavior of GBSA when alternative clustering criteria—namely, the maximization of achievable downlink rate (sumRate) and of channel-norm magnitude (chNorm) are used instead of RANEE. As illustrated, these objectives lead to substantially lower energy-efficiency performance under the same cluster-size constraints. This result further reinforces the suitability of RANEE as an objective function for energy-efficiency–oriented UC clustering.

\begin{figure*}[!htb]
  \begin{minipage}[b]{.40\textwidth}
    \begin{tikzpicture}
    \begin{groupplot}[
        group style={
            group size=2 by 1,       
            horizontal sep=0.2cm,      
            x descriptions at=edge bottom
        },
        width=4.5cm,
        height=5cm,
        xlabel style={font=\scriptsize},
        ylabel style={font=\scriptsize},
        xticklabel style={font=\scriptsize},
        yticklabel style={font=\scriptsize},
        ymin=0,
    ]

    \nextgroupplot[
        xlabel={Normalized Avg RANEE [bit/J]},
        ylabel={Max. UC Cluster Size},
        grid=both,
        legend style={at={(1.00,-0.30)},anchor=north,legend columns=-1,font=\scriptsize},
    ]
        \addplot table [y=MCS, x = arg_max_EE, col sep=comma] {data/compranee.csv};
        \addlegendentry{RANEE}
        \addplot table [y=MCS, x = arg_max_SRk, col sep=comma] {data/compranee.csv};
        \addlegendentry{sumRate}
        \addplot table [y=MCS, x = arg_max_CN, col sep=comma] {data/compranee.csv};
        \addlegendentry{chNorm}
    \nextgroupplot[
        xlabel={Avg Processing Time [s]},  
        grid=both,
        xmin=0.0, xmax=15.0,
        yticklabels={}, 
    ]
        \addplot table [y=MCS, x = arg_max_EE, col sep=comma] {data/compptime.csv};
        \addplot table [y=MCS, x = arg_max_SRk, col sep=comma] {data/compptime.csv};
        \addplot table [y=MCS, x = arg_max_CN, col sep=comma] {data/compptime.csv};      
    \end{groupplot}
    \end{tikzpicture}
    \caption{Performance of GBSA under alternative clustering objectives (sumRate and chNorm) compared to RANEE.}
    \label{fig:plotComp}
  \end{minipage}
    \hspace{7mm}
  \begin{minipage}[b]{.5\textwidth}
    \begin{tikzpicture}
    \begin{groupplot}[
        group style={
            group size=3 by 1,       
            horizontal sep=0.2cm,      
            x descriptions at=edge bottom
        },
        width=4.5cm,
        height=5cm,
        xlabel style={font=\scriptsize},
        ylabel style={font=\scriptsize},
        xticklabel style={font=\scriptsize},
        yticklabel style={font=\scriptsize},
        ymin=0,
    ]
    
    \nextgroupplot[
        xlabel={Normalized Avg RANEE [bit/J]},
        ylabel={Max. UC Cluster Size},
        grid=both,
        legend style={at={(1.50,-0.30)},anchor=north,legend columns=-1,font=\scriptsize},
    ]
    \addplot+[mark=*, thick] table [x=ExSearch, y = MCS, col sep=comma] {data/ranee.csv};
    \addlegendentry{ExSearch}
    
    \addplot+[mark=triangle*, thick] table [x=GBSA, y = MCS, col sep=comma] {data/ranee.csv};
    \addlegendentry{GBSA}
    
    \addplot+[mark=square*, thick] table [x=JO, y = MCS, col sep=comma] {data/ranee.csv};;
    \addlegendentry{JO}
        
    \nextgroupplot[
        xlabel={Avg Processing Time [s]},  
        grid=both,
        yticklabels={}, 
        xmode=log,
    ]
    \addplot+[mark=*, thick] table [x=ExSearch, y = MCS, col sep=comma] {data/proctime.csv};
    
    \addplot+[mark=triangle*, thick] table [x=GBSA, y = MCS, col sep=comma] {data/proctime.csv};
    
    \addplot+[mark=square*, thick] table [x=JO, y = MCS, col sep=comma] {data/proctime.csv};
    
    \nextgroupplot[
        xlabel={Avg UC Cluster Size [APs/UE]}, 
        grid=both,
        xmin=1.0, xmax=4.0,
        yticklabels={}, 
    ]
    \addplot+[mark=*, thick] table [x=ExSearch, y = MCS, col sep=comma] {data/clustersize.csv};
    
    \addplot+[mark=triangle*, thick] table [x=GBSA, y = MCS, col sep=comma] {data/clustersize.csv};
    
    \addplot+[mark=square*, thick] table [x=JO, y = MCS, col sep=comma] {data/clustersize.csv};
        
    \end{groupplot}
    \end{tikzpicture}
    \caption{Comparison of RANEE, processing time, and cluster size across ExSearch, GBSA, and JO}
    \label{fig:plotSNW}
  \end{minipage}
\end{figure*}

Figure \ref{fig:plotSNW} confirms that the exhaustive search consistently reaches the global optimal RANEE, as expected. However, its processing time grows exponentially with the network size, making it infeasible for medium- or large-scale deployments. In contrast, the JO method yields the smallest execution time but produces the lowest RANEE among the evaluated approaches. GBSA offers a favorable compromise: it achieves an RANEE value that closely approaches the exhaustive optimum while maintaining substantially lower computational cost than the exhaustive search and significantly better energy efficiency than JO. Finally, Figure \ref{fig:plotSNW} also shows that, in all evaluated scenarios, the resulting UC cluster size remains smaller than the total number of available APs, indicating that operating the network in a fully cell-free mode is inherently energy-inefficient.

\section{Conclusion}
In this paper, we have addressed the problem of energy-efficient clustering in user-centric cell-free massive MIMO networks by introducing the proposed  algorithm. By modeling the serving states as a Hamming graph and restricting state adjacency exploration to single-bit transitions, GBSA achieves a favorable balance between computational complexity and solution accuracy. Simulation results demonstrate that the proposed approach significantly outperforms conventional hill-climbing methods and attains performance close to that of ExSearch, while requiring substantially fewer computational resources. These findings confirm the potential of graph-based search strategies combined with fractional programming to enable scalable and energy-efficient clustering in next-generation wireless networks. Future work will extend this approach to adaptive neighborhood exploration, larger network dimensions, and integration with machine learning-based heuristics to further enhance scalability and robustness.

\section*{Acknowledgment}
During the preparation of this manuscript, the authors used ChatGPT  for refining academic language. The authors reviewed and revised the material generated and takes full responsibility for the content of this publication.
\bibliographystyle{IEEEtran}
\bibliography{IEEEabrv,ref}
\end{document}